\documentclass[aps,twocolumn,amsmath,amssymb,prl,showpacs]{revtex4}
\usepackage{amsfonts, hyperref}
\usepackage{graphicx}
\begin{document}

\title{Minimization of Ohmic losses for domain wall motion in a
  ferromagnetic nanowire}

\author{O.~A.~Tretiakov}
\author{Y.~Liu}
\author{Ar.~Abanov}

\affiliation{
            Department of Physics,
            MS 4242,
	    Texas A\&M University,
            College Station, TX 77843-4242, USA
}

\date{June 3, 2010}

\begin{abstract}
We study current-induced domain-wall motion in a narrow ferromagnetic
wire.  We propose a way to move domain walls with a resonant
time-dependent current which dramatically decreases the Ohmic losses
in the wire and allows to drive the domain wall with higher speed
without burning the wire.  For any domain wall velocity we find the
time-dependence of the current needed to minimize the Ohmic losses.
Below a critical domain-wall velocity specified by the parameters of
the wire the minimal Ohmic losses are achieved by dc current.
Furthermore, we identify the wire parameters for which the losses
reduction from its dc value is the most dramatic.
\end{abstract}

\pacs{75.78.Fg; 75.60.Ch; 85.75.-d}

\maketitle

\textit{Introduction.}  In recent years there has been intense
interest in applications of domain wall (DW) motion in ferromagnetic
nanowires \cite{Parkin:racetrack08, Allwood02}. This interest is
mostly based on the possibility to store and exchange information by
means of moving domain walls which separate the regions of
magnetization parallel and anti-parallel to the wire. These regions
with parallel and anti-parallel magnetization can be thought of as two
bits, zero and one, of binary information storage.

DWs can be moved by a magnetic field \cite{Ono99, Allwood02} or
electric current \cite{Yamaguchi04, Parkin:racetrack08}. For
technological applications the current driving is preferred as
magnetic field is difficult to apply locally to small wires. Thus, in
this Letter we consider the current-driven DW devices. To achieve
their highest performance it is important to minimize the losses on
Joule heating in the wire, which are due to the resistance of the wire
itself and the entire circuit. They are proportional to the
time-averaged current square, $\langle J^2 \rangle$. Their
minimization has a twofold advantage. First, one can increase the
maximum current which still does not destroy the wire by excessive
heating and therefore move the DWs with a higher velocity, since the
DW velocity increases with the applied current.  Second, it creates
the most energy efficient memory devices and also increases their
reliability.

\begin{figure}
\includegraphics[width=1\columnwidth]{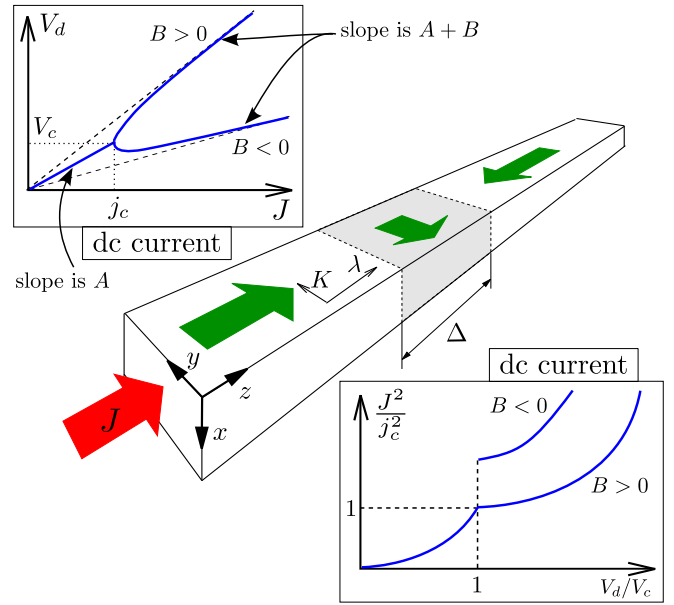}
\caption{(color online) A sketch of a current driven domain wall in
  the ferromagnetic wire. The upper inset shows the dependence of
  drift velocity $V_d$ of DW on dc current $J$ for $B>0$ and $B<0$,
  see Eq.~\eqref{eq:z0Dot}. The slope at $J<j_c$ is given by $A$,
  while at $J\gg j_c$ it is $A+B$. The lower inset shows the power of
  Ohmic losses $p_{\rm{dc}}(V_d/V_c)=J^2/j_c^2$ for dc current. For
  $B<0$ the power has a discontinuity at $V_d/V_c =1$.}
\label{fig:domain_wall} 
\end{figure}

\begin{figure}
\includegraphics[width=1\columnwidth]{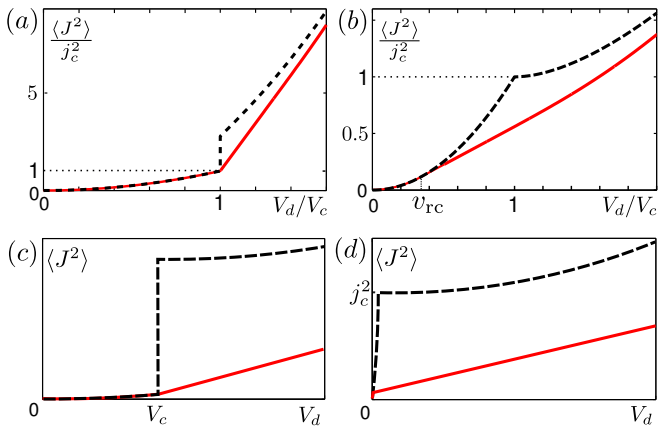} 
\caption{(color online) Minimal power of Ohmic losses
  $\overline{p}=\langle J^2 \rangle /j^2_c$ as a function of drift
  velocity $V_d$ shown by solid line for (a) $a=0.5$ (b) $a=2$.  The
  dashed line depicts $\overline{p}$ for dc current. A sketch of
  $\langle J^2 \rangle (V_d)$ shown by solid line in (c) for $\beta\gg
  \alpha$ ($a\ll 1$) and (d) for $\beta\ll \alpha$ ($a\gg 1$).}
\label{fig:Power_v} 
\end{figure}

To achieve these goals we propose to utilize a ``resonant''
time-dependent current, which allows to gain a significant reduction
of Ohmic losses.  We show that all thin wires can be characterized by
three parameters obtained from dc-driven DW motion experiments:
critical current $j_{c}$, drift velocity at the critical current
$V_{c}$, and material dependent parameter $a>0$, which in particular
depends on Gilbert damping $\alpha$ and non-adiabatic spin torque
constant $\beta$. The parameter $a$ is just a ratio of the slopes of
the drift velocity $V_{d}(J)$ at large and small dc-currents, see the
upper inset of Fig.~\ref{fig:domain_wall}. Our main results are
summarized in Fig.~\ref{fig:Power_v}. We find the minimal power
$\langle J^2 \rangle$ needed to drive a DW with drift velocity $V_d$.
Figures~\ref{fig:Power_v} (a) and (b) show the dependence of power
$\langle J^2 \rangle$ on $V_d$ for the optimal time-dependent current
-- red solid curves, and for dc current -- black dashed curves, for
two cases: (a) $a<1$ and (b) $a>1$. In Fig.~\ref{fig:Power_v} (a) the
minimal power is given by dc current for $V_d <V_{c}$, but above
$V_{c}$ there is a significant reduction in the heating power compared
to dc current. Fig.~\ref{fig:Power_v} (b) shows that the power
$\langle J^2 \rangle$ is reduced in comparison with the dc case for
$V_d >V_c v_{\rm{rc}}$. The (dimensionless) resonant critical velocity
$v_{\rm{rc}}\leq 1$ and can be extracted from the dc-current
measurements.  For Permalloy using \cite{Moriya08} we estimated $a
\approx 0.5$, see Fig.~\ref{fig:Power_v} (a), where for $V_d \gtrsim
V_{c}$ the power is less than $50\%$ of that for the dc current.

Figs.~\ref{fig:Power_v} (c) and (d) show the limiting cases of $a\ll
1$ (c) and $a\gg 1$ (d). We note that for small $\alpha $ and $\beta
$, $a\approx \alpha/\beta$.  If $a\ll 1$ ($\beta\gg\alpha$),
Fig.~\ref{fig:Power_v} (c), for dc current the excessive heating power
$\sim 1/a^2$ essentially limits the highest achievable drift velocity
$V_d$ by $V_c$, whereas the resonant ac-current can move DWs with much
higher $V_d$ (and still rather low power).  In the opposite case $a\gg
1$, ($\beta \ll \alpha$), Fig.~\ref{fig:Power_v} (d), the power saving
starts to be considerable at very small velocity $V_{d}$.  If $\beta
=0$ the dc-current power is finite even at $V_{d}\to 0$, while for the
resonant ac-current the power linearly approaches zero at small $V_d$.
Therefore, our approach gives a dramatic power reduction even in the
least favorable cases $\beta \ll \alpha$ and $\beta \gg \alpha$, thus
opening new doors for using materials with much wider range of $\beta$
for fast DW motion.

\textit{Model.}  DW in a ferromagnetic wire can be modeled by a
Hamiltonian which contains exchange and dipolar interactions. In a
thin wire, the latter can be approximated by two anisotropies: along
the wire ($\lambda$) and transverse to it ($K$). A sketch of a wire
with a DW of width $\Delta$ is shown in
Fig.~\ref{fig:domain_wall}. The dynamics of magnetization $\mathbf{S}$
in a wire is described by Landau-Lifshitz-Gilbert (LLG) equation with
the current $J$ \cite{Zhang04, Thiaville05},
\begin{equation}
  \dot{\mathbf{S}}=\mathbf{S}\times\mathbf{H}_{e}
-J\partial\mathbf{S}
+\beta J\mathbf{S}\times\partial\mathbf{S}
+\alpha\mathbf{S}\times\dot{\mathbf{S}},
\label{eq:LLG}
\end{equation}
where $\mathbf{H}_{e}=\delta\mathcal{H}/\delta\mathbf{S}$ is the
effective magnetic field given by the Hamiltonian $\mathcal{H}$ of the
system, $\alpha$ is Gilbert damping constant, $\beta$ is non-adiabatic
spin torque constant, and $\partial=\partial/\partial z$.
Furthermore, it can be shown~\cite{Tretiakov_DMI} that in a thin wire
the DW is a rigid spin texture for not too strong applied currents and
its dynamics can be described in terms of only two collective
coordinates (corresponding to the two softest modes of the DW motion),
namely, the position of the DW along the wire $z_0$ and the rotation
angle $\phi$ of the magnetization in the DW around the wire axis.

To describe the DW dynamics we need to find the equations of
motion. For the two softest modes of the DW, $z_0(t)$ and $\phi (t)$,
they can be found as an expansion in small current $J$ up to a linear
in $J$ order. Due to the translational invariance $\dot{z}_{0}$ and
$\dot{\phi}$ cannot depend on $z_0$. In addition, to the first order
in small transverse anisotropy $K$, $\dot{\phi}$ and $\dot{z}_{0}$ are
proportional to the first harmonic $\sin(2\phi)$. Then the most
general equations of DW motion are
\begin{subequations}
\label{eq:z0Dot_phiDot}
\begin{eqnarray} 
& & \dot{\phi}=C[J-j_{c}\sin(2\phi)]\label{eq:phiDot},\\ 
& & \dot{z}_{0}=AJ+B[J-j_{c}\sin(2\phi)],
\label{eq:z0Dot}
\end{eqnarray}
\end{subequations}
where $J(t)$ is, in general, a time-dependent current whose frequency
is not too high to create spin waves and other excitations in the
wire. Coefficients $A$, $B$, $C$, and critical current $j_c$ can be
calculated for a particular model \footnote{As it was shown in
  Ref.~\cite{Tretiakov_DMI}, $A= \beta/\alpha$, $B=
  (\alpha-\beta)(1+\alpha\Gamma\Delta)/[\alpha(1+\alpha^{2})]$, $C=
  (\alpha-\beta) \Delta/[(1+\alpha^{2})\Delta_{0}^{2}]$, and
  $j_c=(\alpha K\Delta /\left|\alpha-\beta\right|) [\pi \Gamma\Delta
    /\sinh(\pi\Gamma\Delta)]$. Here $\Gamma =D/J_{\rm{ex}}$,
  $J_{\rm{ex}}$ is exchange constant, and $D$ is Dzyaloshinskii-Moriya
  interaction (DMI) constant. Also, $\Delta=\Delta_0 /\sqrt{1-\Gamma^2
    \Delta_0^2}$ where $\Delta_0$ is the DW width in the absence of
  DMI.} in terms of $\alpha$, $\beta$ and other microscopic parameters
by means of deriving Eqs.~\eqref{eq:z0Dot_phiDot} from LLG
equation~\eqref{eq:LLG}. However, we emphasize that
Eqs. \eqref{eq:z0Dot_phiDot}, with coefficients $A$, $B$, $C$, and
$j_c$ determined directly from dc-current experiment for each
particular wire, have more general validity than just being derived
from LLG, e.g., due to the complicated influence of disorder and
internal DW dynamics \cite{Min10}. Namely, the value of $j_c$ is
defined as the endpoint of the linear regime of the time-averaged
(drift) velocity $V_d =\langle \dot{z}_{0}(J) \rangle$, see the upper
inset of Fig.~\ref{fig:domain_wall}. The linear slope of $V_d(J)$
below $j_c$ determines constant $A$. The slope of $V_d(J)$ at large
$J$ gives $A+B$. Constant $C$ one can obtain, e.g., from the
measurements of the DW electromotive force~\cite{BeachPRL09, LiuEMF}
for dc current.

\textit{DC current.}  For the dc current applied to the wire the DW
dynamics governed by Eqs.~\eqref{eq:z0Dot_phiDot} can be obtained
explicitly~\cite{Tretiakov_DMI}. For $J<j_c$ and $A\neq 0$ the DW
moves along the wire but does not rotate around its axis. It only
tilts on angle $\phi_0$ from the transverse-anisotropy easy axis ($y$
axis) given by condition $\sin(2\phi_0)=J/j_c$. The drift velocity is
given by $V_{d}=AJ$, see Eq.~\eqref{eq:z0Dot}. At $J=j_{c}$ the
magnetization angle becomes perpendicular to the easy axis,
$\phi_0=\pi/2$.  For $J>j_{c}$ the DW both moves and rotates, and
Eqs.~\eqref{eq:z0Dot_phiDot} give $V_{d} = AJ
+B\sqrt{J^{2}-j_{c}^{2}}$~\cite{Tretiakov_DMI}.

The influence of the spin structure on the current is negligible.  The
largest losses in the system are the Ohmic losses of the current.  The
power of Ohmic losses is proportional to $J^2$.  Therefore, at $J<j_c$
the current is $J=V_{d}/A$ and the power of Ohmic losses is
$\mathcal{P}_{\rm{dc}}=J^{2}=V_{d}^{2}/A^{2}$. It is instructive to
introduce the dimensionless variables for time, drift velocity,
current, and power. Using $V_{c}=Aj_{c} \simeq K\Delta$ we
find~\footnote{It can be shown that $C\sim B \sim
  \alpha-\beta$~\cite{Tretiakov_DMI}. In the special case of $\alpha
  =\beta$, we find $C=B=0$ and one cannot use dimensionless
  variables~\eqref{eq:dimensionless}. The DW dynamics in this case is
  trivial \cite{Barnes05}. The DW does not rotate $\dot{\phi}=0$ and
  moves with the velocity given by current $\dot{z}_0=J$.}
\begin{equation}
\tau=Cj_{c}t,\quad
v_{d}=V_{d}/V_{c},\quad j=J/j_c,\quad
p=\mathcal{P}/j_{c}^{2}.
\label{eq:dimensionless}
\end{equation} 
Using Eq.~\eqref{eq:dimensionless} we find $p_{\rm{dc}}=v_{d}^{2}$ for
$v_{d}<1$.

For currents above $j_c$ the dimensionless power $p_{\rm{dc}}$ is
given in terms of dimensionless drift velocity
$v_{d}=j+(B/A)\sqrt{j^{2}-1}$ as $p_{\rm{dc}}(v_{d})=j^2$, see the
lower inset of Fig.~\ref{fig:domain_wall}. Thus, it is quadratic in
$v_d$, and at $v_{d}\gg 1$ it approaches $p_{\rm{dc}}
=A^2v_{d}^{2}/(B+A)^{2} +B/(B+A)$.  For $B>0$ right above $v_{d}=1$,
it is approximated by $p_{\rm{dc}} =1+(A/B)^{2}(v_d -1)^{2}$. For
$B<0$ the power has a discontinuity at $v_d =1$.

\textit{Minimization of Ohmic losses by time-dependent current.}  In
this part we minimize the Ohmic losses while keeping the DW moving
with a given drift (average) velocity. Equations of
motion~\eqref{eq:z0Dot_phiDot} are correct even when the current
depends on time.  In general, the DW motion has some period $T$ and
current $j(\tau)$ must be a periodic function with the same $T$ to
minimize the Ohmic losses.

In the following it is more convenient to measure the angle from the
hard axis instead of easy axis and to scale it by factor of 2, so that
$2\phi= \theta -\pi/2$. Also, we introduce the ratio of slopes of
$V_d(J)$ at large and small currents $a=(A+B)/A$. Then using
Eq.~\eqref{eq:phiDot} the dimensionless current becomes
\begin{equation}
j(\tau)= \dot{\theta}/2 -\cos\theta,
\label{eq:j}
\end{equation}
where $\dot{\theta}=\partial\theta/\partial\tau$.  Averaging
Eq.~\eqref{eq:z0Dot} over dimensionless period $T$ we find
\begin{equation}
v_{d}=\frac{a}{2}\langle\dot{\theta}\rangle
-\langle\cos\theta\rangle ,
\label{eq:VdAve}
\end{equation}
where $\langle\dots\rangle = \frac{1}{T} \int_{0}^{T} \dots d\tau$ is
the time averaging.

To minimize the power of Ohmic losses $\overline{p}$ averaged over
time we need to find the minimum of $\langle j^{2}(\tau) \rangle$ at
fixed $v_{d}$ given by Eq.~\eqref{eq:VdAve},
\begin{equation}
\overline{p} = \left\langle
\left(\frac{\dot{\theta}}{2}-\cos\theta\right)^{2}
-2\rho \left(\frac{a}{2}\dot{\theta}
-\cos\theta -v_{d}\right)\right\rangle . 
\label{eq:Power}
\end{equation}
Here to account for the constraint given by Eq.~\eqref{eq:VdAve} we
used a Lagrange multiplier $2\rho$, with $\rho$ being an arbitrary
dimensionless constant. Note that the cross term $\sim\int
\dot{\theta}\cos\theta\, d\tau'$ and the term $\sim\int \dot{\theta}\,
d\tau'$ can be dropped for the minimization procedure as they are full
derivatives.

Power~\eqref{eq:Power} can be considered as an effective action for a
hypothetical particle of mass $1/2$ in a periodic potential field, and
its minimization leads to the equation of motion
\begin{equation}
\frac{\ddot{\theta}}{2}
=-\frac{\partial U}{\partial \theta},
\qquad U(\theta,\rho)=-\cos^{2}\theta-2\rho\cos\theta.
\label{eq:EOM}
\end{equation}
It can be reduced to the first order differential equation
\begin{equation}
\dot{\theta} = \pm 2\sqrt{d-U(\theta,\rho)},
\label{eq:thetaDotD}
\end{equation}
where $d$ is an arbitrary integration constant. Note that changing
$\rho\to -\rho$ in $U$ of Eq.~\eqref{eq:EOM} is equivalent to changing
$\theta\to \pi+\theta$, so below we consider only positive $\rho$. The
potential has a minimum at $\theta =0$ with $U_{\rm{min}}=2\rho -1$
for any $\rho\geq 0$. For $\rho <1$ it has also minimum at
$\theta=\pm\pi$ with $U(\pm\pi)=-2\rho -1$ and the maximum at
$\cos\theta_\rho = -\rho$ with $U(\pm\theta_\rho)=\rho^2$. For $\rho
>1$ it has maximum at $\theta=\pm\pi$ with $U(\pm\pi)=2\rho -1$.

\begin{figure}
\includegraphics[width=0.9\columnwidth]{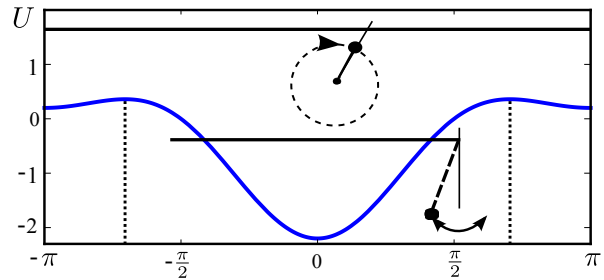} 
\caption{(color online) Potential $U(\theta)$ in which a ``particle''
  is moving in the rocking (pendulum-like) and rotational regimes.}
\label{fig:Potential} 
\end{figure}

According to Eq.~\eqref{eq:thetaDotD} there are two different regimes:
i) the rocking regime where $d< \rm{max}[U(\theta,\rho)]$ in which
case $\theta$ is bounded, and the particle oscillates in potential
well $U(\theta)$, see Fig.~\ref{fig:Potential}; and ii) the rotational
regime where $d> \rm{max}[U(\theta,\rho)]$ in which case the
magnetization in the DW rotates. Below we consider these regimes
separately.

\textit{Rocking regime.} In this regime the motion of $\theta$ mimics
pendulum motion.  The particle rocks between the two turning points
$-\theta_0$ and $\theta_0$ given by the condition $d=U(\pm
\theta_0,\rho)$.  At these points $\dot{\theta}=0$. Since $\theta$ is
a bounded function $\langle\dot{\theta}\rangle=0$ and the averaged
velocity becomes $v_{d} =-\langle\cos\theta\rangle$.  The averaging is
done over a period of one complete oscillation,
\begin{equation}
T=\int_{-\theta_{0}}^{\theta_{0}}\frac{d\theta}{\sqrt{d-U(\theta ,\rho )}},
\label{eq:tau}
\end{equation}
and according to Eq.~\eqref{eq:Power} the power is given by
\begin{equation}
\overline{p}=  
\langle {\dot{\theta}}^2 \rangle /4
+\langle\cos^{2} \theta \rangle .
\label{eq:p}
\end{equation}

Most generally $\theta$ depends on time.  For any $\theta(\tau)$,
however, $\langle \cos^{2}\theta \rangle \geq \langle \cos\theta
\rangle^{2}$.  Then from Eq.~\eqref{eq:p} follows $\overline{p}\geq
\langle \dot{\theta}^{2}\rangle/4 +\langle \cos \theta \rangle^{2} =
\langle \dot{\theta}^{2}\rangle/4 +v_{d}^{2} \geq
v_{d}^{2}=p_{\rm{dc}}$. Thus, in the bounded regime the power of Ohmic
losses is minimal for dc current and is given by $\overline{p}=v_d^2$.

\textit{Rotational regime.}  Next we study the case when $d>
\rm{max}[U(\theta,\rho)]$, so that angle $\theta$ is unbounded. It
corresponds to the rotational motion of the transverse to the wire
component of the DW magnetization. Note that in the rotational regime
the term in Eq.~\eqref{eq:VdAve} with $\langle\dot{\theta}\rangle$
should be kept because $\theta$ is not bounded. The time it takes for
$\theta$ to make a full rotation from $-\pi $ to $\pi $ defines the
period $T$.  Then the period, drift velocity, and power, according to
Eq.~\eqref{eq:p}, are given by
\begin{subequations}
\label{eq:rotational}
\begin{eqnarray}
T & = & \frac{1}{2}\int_{-\pi}^{\pi}
\frac{d\theta}{\sqrt{d-U(\theta,\rho)}},
\label{eq:tauUn}\\
v_{d} & = & \frac{\pi a}{T}
-\frac{1}{2T}\int_{-\pi}^{\pi}\!\!
\frac{\cos\theta\, d\theta}{\sqrt{d-U(\theta,\rho)}},
\label{eq:vdUn}\\
\overline{p} & = & 
\frac{1}{2T}\int_{-\pi}^{\pi}\!\!
\frac{d-U(\theta,\rho)+\cos^2 \theta}{\sqrt{d-U(\theta,\rho)}}d\theta.
\label{eq:pUn}
\end{eqnarray}
\end{subequations}
This system of equations, after minimizing the power $\overline{p}$
with respect to both $d$ and $\rho$ at fixed $v_d$, gives
$\overline{p}(v_d)$. One can either directly perform a numerical
minimization of Eq.~\eqref{eq:rotational} or alternatively try to find
the minimization condition for $\overline{p}$ analytically. We have
followed both routes.

The minimization of Eqs.~\eqref{eq:rotational} infers that $\partial
\overline{p}/\partial \rho |_{v_d}=0$ from which one can
find~\cite{Ohmic_suppl_material}
\begin{equation}
\int_{-\pi}^{\pi } \!\sqrt{d-U(\theta ,\rho )}d\theta =2\pi a\rho. 
\label{eq:min_condition}
\end{equation}
This equation gives the relationship between $d$ and $\rho$. Solving
it together with Eq.~\eqref{eq:vdUn} one finds $d$ and $\rho$ in terms
of $v_d$. They are then substituted into $\overline{p} =2\rho v_d -d$
which follows from Eqs.~\eqref{eq:pUn} and~\eqref{eq:min_condition}.
The motion is unbounded when $d> \rm{max}[U(\theta,\rho)]$, see
Eq.~\eqref{eq:thetaDotD}, which leads to $d>\rho^2$ for $\rho<1$ and
$d>2\rho-1$ for $\rho>1$.

The results for the minimal power of Ohmic losses $\overline{p}(v_d)$
are presented in Fig.~\ref{fig:Power_v}.  For $a<1$, see
e.g. Fig.~\ref{fig:Power_v} (a), at $v_d<v_{\rm{rc}}=1$ the minimal power
$\overline{p}$ coincides with the one given by dc current, whereas at
$v_d>1$ it is significantly lower than $p_{\rm{dc}}$. Immediately
above $v_d=1$ we find that there is a range of $v_d$ where
$\overline{p}=1+2\rho_0 (v_d -1)$ with $\rho_0(a)>1$ given by
Eq.~\eqref{eq:min_condition} with $d= 2\rho -1$. Therefore,
$\overline{p}$ is linear in $v_d$ right above $v_{\rm{rc}}=1$.

For $a>1$, see e.g. Fig.~\ref{fig:Power_v} (b), we
show~\cite{Ohmic_suppl_material} that there is a critical velocity
$v_{\rm{rc}}<1$, such that at $v_d<v_{\rm{rc}}$ the power of Ohmic
losses is $\overline{p}=v_d^2=p_{\rm{dc}}$. Above $v_{\rm{rc}}$ one
can minimize the Ohmic losses by moving DW with resonant current
pulses. Right above $v_{\rm{rc}}$ there is a certain range of $v_d$
where $d\simeq \rho^2$, and therefore we find $\overline{p}=2\rho_0
v_d -\rho_0^2$ with $\rho_0(a)<1$ given by
Eq.~\eqref{eq:min_condition} with $d= \rho^2$. The critical velocity
is found as $v_{\rm{rc}}=\rho_0(a)$.  For $a\gg 1$ (corresponding to
non-adiabatic spin transfer torque coefficient $\beta\ll \alpha$,
c.f. Eq.~\eqref{eq:LLG}) we find $v_{\rm{rc}} \simeq 2/(\pi a)$ and
therefore for $v_d\gg v_{\rm{rc}}$ we obtain $\overline{p}=4v_d/(\pi
a)$.

We show that at large $v_d$ the minimal power is always smaller than
$p_{\rm{dc}}$. Note that for $d\gg 1$ Eq.~\eqref{eq:min_condition}
gives $d=a^2\rho^2 $.  Using it we find that the difference between
them approaches $p_{\rm{dc}}-\overline{p} =(1-1/a)^2/2$ at $v_d \gg
1$.

\begin{figure}
\includegraphics[width=1\columnwidth]{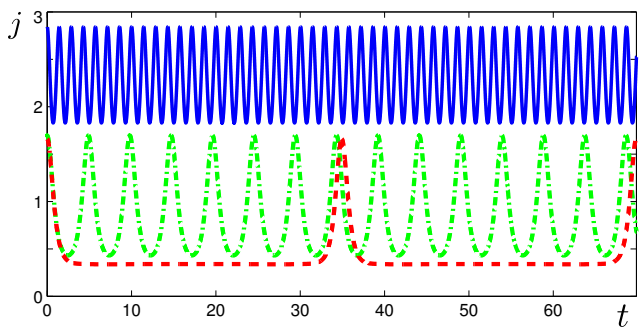} 
\caption{(color online) Current $j$ as a function of time $\tau$ at
  velocities $v_d =0.5$ (dashed line), $v_d =1.5$ (dot-dashed line),
  and $v_d =4.5$ (solid line) in the rotational regime for $a=2$.}
\label{fig:currents_rotation} 
\end{figure}

\textit{Optimal current.} For $v_d<v_{\rm{rc}}$ the optimal current
coincides with the dc current. Above $v_{\rm{rc}}$ the resonant
current $j(t)$ is plotted in Fig.~\ref{fig:currents_rotation} for
different velocities $v_d$ in the case $a=2$. At small $v_d$ the
current is given by $j(\tau)= -2\cos(\theta (\tau))-v_{\rm{rc}}$ for
$\cos(\theta (\tau))<-v_{\rm{rc}}$ and by $j(\tau)=v_{\rm{rc}}$ for
$\cos(\theta (\tau))> -v_{\rm{rc}}$. At $v_d\gg 1$ the current is
approximated by $j\approx v_d/a +[(1-a)/a]\cos\theta$.

In general, at $v_d>v_{\rm{rc}}$ the current's maximum $j_{\rm{max}}$
increases from $2-v_{\rm{rc}}$ at small enough $v_d\lesssim 1$ up to
$j_{\rm{max}}\approx v_d/a$ at $v_d\gg 1$.  The current's minimum
increases monotonically from small positive values $j_{\rm{min}}=
v_{\rm{rc}}$ at $v_d\sim 1$ up to $j_{\rm{min}}=j_{\rm{max}}
-2|1-a|/a$ at $v_d \gg 1$. At $v_d\lesssim 1$ (for $a>1$) the time
between the current picks decreases with increasing velocity as $T
\simeq (\pi a-2\arcsin v_{\rm{rc}})/(v_d -v_{\rm{rc}})$, whereas the
pick's width is given by $\approx 1.3/\sqrt{(1-v_{\rm{rc}})}$, which
is independent of $v_d$. Therefore, at small $v_d-v_{\rm{rc}}$ the
picks are widely separated, then as $v_d$ increases the time between
the picks decreases. At $v_d \gg 1$ the optimal current has a large
constant component, which is close to but smaller than the dc current
for the same $v_d$, and has small-amplitude ac modulations with a
period $T\approx \pi a/v_d$ on top of it.

\textit{Summary.}  We have studied the current driven DW dynamics in
thin ferromagnetic wires. We have found the ultimate lower bound for
the Ohmic losses in the wire for any DW drift velocity $V_d$. The
explicit time-dependence of current, see
Fig.~\ref{fig:currents_rotation}, has been found which minimizes the
Ohmic losses. We have shown that the use of these specific current
pulses instead of applying dc current can help to significantly reduce
heating of the wire for any $V_d$. Even in the limiting cases of the
systems with weak ($\beta\ll \alpha$) or strong ($\beta\gg \alpha$)
non-adiabatic spin transfer torque, where the power of Ohmic losses is
high for dc currents, the optimized ac current gives significant
reduction in heating power thus greatly expanding the range of
materials which can be used for spintronic devices \cite{Allwood02,
  Parkin:racetrack08}.

We are grateful to J.~Sinova for valuable discussions.  This work was
supported by the NSF Grant No. 0757992 and Welch Foundation (A-1678).

\bibliography{magnetizationDynamics}

\newpage

\section*{Supplementary material for ``Minimization of Ohmic losses for
  domain wall motion in a ferromagnetic nanowire''}

\section{Minimization procedure}\label{sec:intro}

As described in the main part of the Letter we study a domain-wall
dynamics under the influence of a time-dependent current. The
equations of motion for the domain wall in the thin ferromagnetic wire
take the form of Eqs. (2) of the main part of this Letter. To obtain
the results of the main part, we use the dimensionless variables
introduced in Eq. (3) of the main part.

In order to minimize the power of Ohmic losses we need to find a
minimum of the average $j^{2}$,
\begin{equation}\label{eq:Pp}
\overline{p}=\langle j^{2}(\tau)\rangle
\end{equation}
at a fixed drift velocity $v_d$.  From Eqs. (2) of the main part of
the Letter we find
\begin{eqnarray}
&&j(\tau )=\frac{1}{2}\dot{\theta}-\cos \theta,  
\label{eq:jj} \\
&&v(\tau )=\frac{a}{2}\dot{\theta}-\cos \theta, 
\label{eq:v}\\
&&v_{d}=\frac{a}{2}\langle \dot{\theta}\rangle-\langle \cos \theta \rangle,
\label{eq:vd}\\
&&\overline{p}=\left\langle \left( \frac{1}{2}\dot{\theta} 
-\cos \theta\right)^{2}\right\rangle , 
\label{eq:pp}
\end{eqnarray}
where $\dot{\theta}= \partial\theta/\partial\tau$ and the averaging is
performed over the dimensionless period $T$ of the magnetization
oscillations.  To find the minimum of power at fixed drift velocity we
introduce a Lagrange multiplier $2\rho$ and minimize the functional
\begin{equation}\label{eq:A}
\overline{p}=\left\langle \left( \frac{1}{2}\dot{\theta}
-\cos \theta\right)^{2}\right\rangle 
-2\rho \left( \frac{a}{2}\langle \dot{\theta}\rangle
-\langle \cos \theta \rangle-v_{d}\right),
\end{equation}
where $\langle \dots \rangle =T^{-1}\int_{0}^{T}\dots d\tau$ is the
averaging over time.  Then it follows
\begin{equation}\label{eq:A1}
\overline{p}= 
\frac{1}{T}\int_{0}^{T}\left(  \frac{1}{4}\dot{\theta}^{2}
+\cos^{2} \theta+2\rho\cos \theta \right) d\tau,
\end{equation}
where we dropped all the terms which are full derivatives and
therefore do not give any contribution to the minimized power.

The minimization of Eq.~\eqref{eq:A1} gives the following equation of motion
\begin{eqnarray}
&&\frac{1}{2}\ddot{\theta }=-\partial_{\theta}U(\theta ,\rho ),
\label{eq:eqMotion} \\
&&U(\theta,\rho  )=-\cos^{2}\theta -2\rho \cos \theta.
\label{eq:U}
\end{eqnarray}
Its solution is given by
\begin{equation}\label{eq:sol}
\dot{\theta }=\pm 2\sqrt{d-U(\theta, \rho )}
\end{equation}
where $d$ is an arbitrary constant of integration.  We note, that
changing $\rho \rightarrow -\rho $ is equivalent to changing $\theta
\rightarrow \pi +\theta $, so we only need to consider positive $\rho
$.  The extrema of potential $U$ are
\begin{equation}\label{eq:extr}
\begin{array}{lll}
\theta =0, &U(0,\rho )=-1-2\rho, \\
\theta =\pm \pi, &U(\pm \pi,\rho )=-1+2\rho, \\
\cos \theta_{\rho } =-\rho,\quad  &U(\pm \theta_{\rho },\rho)=\rho^{2}, 
\end{array}
\end{equation}
where the last extremum exists only for $\rho <1$.

We see that there are two different cases: i) {\em rocking} -- when $d$
is smaller then the maximum of $U(\theta ,\rho )$, and ii) {\em
  rotating} -- when $d$ is larger then the maximum of $U(\theta ,\rho
)$. We consider them separately.

\section{Rocking regime}\label{sec:rocking}
In this case there is an angle $\theta_{0}$ such that
$d=U(\theta_{0},\rho )$.  Then we have the oscillating motion between
$-\theta_{0}$ and $\theta_{0}$ and back.  The period of one complete
oscillation, as well as the drift velocity and power are given by
\begin{eqnarray}
T&=&\int_{-\theta_{0}}^{\theta_{0}}\frac{d\theta}{\sqrt{U(\theta_{0},\rho )
-U(\theta ,\rho )}},
\label{eq:period}\\
v_{d}&=&-\langle \cos \theta \rangle 
=-\frac{1}{T}\int_{-\theta_{0}}^{\theta_{0}}\frac{\cos \theta\, d\theta}{
\sqrt{U(\theta_{0},\rho )-U(\theta ,\rho)}},
\label{eq:rockV}\\
\overline{p}&=&\left\langle  \frac{1}{4}\dot{\theta}^{2}
+\cos^{2} \theta\right\rangle
\nonumber\\
&=&\frac{1}{T}\int_{-\theta_{0}}^{\theta_{0}}\frac{U(\theta_{0},\rho )-U(\theta ,\rho )
+\cos^{2}\theta}{\sqrt{U(\theta_{0},\rho )-U(\theta ,\rho )}}d\theta.
\label{eq:rockP}
\end{eqnarray}

\subsection{dc current}\label{sec:rockDC}

In the case of dc current $\theta =\mbox{const}$ equation
\eqref{eq:sol} gives $\theta =\theta_{0}$, then \eqref{eq:eqMotion}
requires that $\theta_{0}$ is an extremum of $U(\theta ,\rho )$, there
are three possibilities $\theta =\{0,\pi ,\theta_{\rho } \}$, for them
we get $v_{d}=\{-1,1,\rho \}=j$, and $p=\{1,1,\rho^{2} \}$. So we see,
that $p=v_{d}^{2}$.

\subsection{ac current}\label{sec:rockAC}

Now we take $\theta$ changing with time. From the condition 
\begin{equation}
\left\langle \left(\langle \cos \theta \rangle-\cos \theta  \right)^{2} 
\right\rangle \geq 0 ,
\end{equation}
it follows that
\begin{equation}
\langle \cos^{2} \theta \rangle \geq \langle \cos \theta \rangle^{2}.
\end{equation}
Then the power $\overline{p}$ satisfies the condition
\begin{eqnarray}\label{eq:rockSol}
\overline{p}&=& \frac{1}{4}\left\langle \dot{\theta}^{2}\right\rangle
+\left\langle \cos^{2} \theta\right\rangle\geq
\frac{1}{4}\left\langle \dot{\theta}^{2}\right\rangle
+\left\langle \cos \theta\right\rangle^{2}
\nonumber\\
&=&\frac{1}{4}\left\langle \dot{\theta}^{2}\right\rangle
+v_{d}^{2}\geq v_{d}^{2}
\end{eqnarray}
Thus, we see that in the rocking regime the dc current minimizes the
Ohmic losses.

\section{Rotating regime}\label{sec:Rotating}

In this section we study the rotational regime. In this case $\theta $
makes a full rotation from $-\pi $ to $\pi $.  The period of one
complete rotation, as well as the drift velocity and power are given
by Eqs. (11) in the main part of this Letter, i.e.
\begin{eqnarray}
T\!&=\!&\frac{1}{2}\int_{-\pi }^{\pi }\frac{d\theta}{\sqrt{d-U(\theta ,\rho )}},
\label{eq:rotT}\\
v_{d}\!&=\!&\frac{a}{2}\langle \dot{\theta }\rangle-\langle \cos \theta \rangle 
=\frac{\pi a}{T}-\frac{1}{2T}\int_{-\pi }^{\pi }\!
\frac{\cos \theta\,d\theta}{\sqrt{d-U(\theta ,\rho )}},
\label{eq:rotV}\\
\overline{p}\! &=\!&\left\langle  \frac{1}{4}\dot{\theta}^{2}
+\cos^{2} \theta \right\rangle
=\frac{1}{2T}\int_{-\pi }^{\pi }\frac{d-U(\theta ,\rho )
+\cos^{2}\theta}{\sqrt{d-U(\theta ,\rho )}}d\theta,
\nonumber\\ 
\label{eq:rotP}
\end{eqnarray}
where $a=(A+B)/A$.

\subsection{dc current}\label{sec:rotDC}

In this case $j=\mbox{const}$ and according to Eq.~\eqref{eq:jj} we find
\begin{eqnarray}
\label{eq:dtau}
&&\frac{1}{2}\dot{\theta }-\cos \theta =j,\quad 
T=\int_{0}^{2\pi }\!\frac{d\theta}{j+\cos \theta} 
= \frac{\pi }{\sqrt{j^{2}-1}},
\nonumber \\
&&\langle \cos \theta \rangle =-j+\frac{1}{2}\langle 
\dot{\theta}\rangle,\quad v_{d}
=\frac{a-1}{2}\langle \dot{\theta }\rangle +j. 
\end{eqnarray}
Since $\langle \dot{\theta }\rangle=2\pi /T$ we obtain
\begin{equation}\label{eq:rotDC}
v_{d}=\frac{a-1}{2}\sqrt{j^{2}-1}+j,\quad p_{\rm{dc}}=j^{2},\quad j>1.
\end{equation}
In particular for $j\gg 1$ we find
\begin{equation}\label{eq:rotDCgg1}
p_{\rm{dc}}\approx \frac{4}{(1+a)^{2}}v_{d}^{2}+\frac{a-1}{a+1}.
\end{equation}

\subsection{ac current}\label{sec:rotAC}

The same trick as in Sec.~\ref{sec:rockAC} gives
\begin{equation*}
\overline{p}\geq \frac{v_{d}^{2}}{1+a^{2}}+\left(\frac{\sqrt{1+a^{2}}}{2}\langle \dot{\theta}\rangle-\frac{a}{\sqrt{1+a^{2}}}v_{d} \right)^{2},
\end{equation*}
and we can only conclude that the power is no smaller than quadratic
in $v_d$.

\subsubsection{Derivation of minimization condition, Eq.~(12)}\label{sec:v1Un}

In this section we derive the minimization condition, Eq.~(12) of the
main part of the Letter.  We rewrite Eqs.~(11) of the main part in the
following way
\begin{eqnarray}
\!\!&&T =\frac{1}{2} \int_{-\pi}^{\pi}\frac{d\theta}{\sqrt{d-U(\theta ,\rho )}}
\label{eq:tauUnV1} \\
\!\!&&T v_d =\pi a - \frac{1}{2}\int_{-\pi}^{\pi}\frac{\cos\theta\, d\theta }{
\sqrt{d-U(\theta ,\rho )}},
\label{eq:vdUnV1}\\
\!\!&&\overline{p}=
2\rho v_d -d
+\frac{1}{T}\int_{-\pi}^{\pi} \left[\sqrt{d-U(\theta ,\rho )}-a\rho  \right]\! d\theta.
\label{eq:pUnV1}
\end{eqnarray}

The minimization of $\overline{p}$ means that $\partial
\overline{p}/\partial \rho |_{v_d}=0$, and we find
\begin{equation}
\left.\frac{\partial \overline{p}}{\partial \rho } \right|_{v_d}=
  -\left.\frac{\partial T }{\partial \rho }
  \right|_{v_d}\frac{1}{T^{2}}\int_{-\pi}^{\pi}
  \left[\sqrt{d-U(\theta ,\rho )}-a\rho \right] d\theta =0
\end{equation}
There are two possibilities to satisfy this condition
\begin{eqnarray}
&&\left.\frac{\partial T }{\partial \rho } \right|_{v_d}=0,\quad \mbox{or} 
\label{eq:pos1}\\
&&\int_{-\pi}^{\pi}\left[\sqrt{d-U(\theta ,\rho )}-a\rho  \right] d\theta=0.
\label{eq:pos2}
\end{eqnarray}
Note that there is also a possibility $T =\infty $ but it corresponds
to a dc-current case.

First we consider the possibility given by
Eq.~\eqref{eq:pos1}. Differentiating Eqs.~\eqref{eq:tauUnV1} and
\eqref{eq:vdUnV1} with respect to $\rho$ at fixed $v_d$ and using
Eq.~\eqref{eq:pos1}, we obtain
\begin{eqnarray}
0\!\!&=\!\!&\left.\frac{\partial d}{\partial \rho } \right|_{v_d}
\int_{-\pi}^{\pi} 
\frac{d\theta}{[d-U(\theta ,\rho )]^{3/2}} 
+2\!\int_{-\pi}^{\pi} \frac{\cos \theta\, d\theta}{[d-U(\theta ,\rho )]^{3/2}} 
\nonumber\\
0\!\!&=\!\!&\left.\frac{\partial d}{\partial \rho } \right|_{v_d}
\int_{-\pi}^{\pi} \frac{\cos \theta\, d\theta}{[d-U(\theta ,\rho )]^{3/2}} 
+2\!\int_{-\pi}^{\pi} \frac{\cos^{2} \theta\,d\theta}{[d-U(\theta ,\rho )]^{3/2}}
\nonumber
\end{eqnarray}
Combining these two equations, we find
\begin{widetext}
\begin{equation}\label{eq:case1}
\left[\int_{-\pi}^{\pi}\frac{\cos\theta\, d\theta}{[d-U(\theta ,\rho )]^{3/2}} 
\right]^{2}
=\int_{-\pi}^{\pi} \frac{d\theta }{[d-U(\theta ,\rho )]^{3/2}} 
\int_{-\pi}^{\pi} \frac{\cos^{2}\theta' \, d\theta'}{[d-U(\theta' ,\rho )]^{3/2}} 
\end{equation}
\end{widetext}
Note that this is a standard Bunyakovsky (Cauchy -- Schwarz) inequality
and it never becomes an equality except for $\theta =0$. Therefore, we
conclude that in the rotational regime the minimization condition is
given by Eq.~(\ref{eq:pos2}).  Then, the system of equations which we
need to solve is
\begin{eqnarray}
&& \int_{-\pi}^{\pi} \left[\sqrt{d-U(\theta ,\rho )}-a\rho  \right] d\theta=0,
\label{eq:1}\\
&&v_d T =\pi a - \frac{1}{2} \int_{-\pi}^{\pi} 
\frac{\cos\theta\, d\theta }{\sqrt{d-U(\theta ,\rho )}},
\label{eq:2}\\
&&\overline{p}=2\rho v_d -d,\label{eq:3}
\end{eqnarray}
where the second equation is just Eq.~\eqref{eq:vdUnV1} with $T$ given
by Eq.~\eqref{eq:tauUnV1}. Eq.~\eqref{eq:1} provides the
correspondence between parameters $\rho$ and $d$. Solving together the
system of equations~\eqref{eq:1} and \eqref{eq:2} yields $\rho (v_d)$
and $d(v_d)$. They are then substituted into Eq.~\eqref{eq:3} to find
$\overline{p}(v_d)$. In general, the system of equations~\eqref{eq:1}
and \eqref{eq:2} has to be solved numerically. However, in the
limiting cases of small and large drift velocities $v_d$ the result
for $\overline{p}(v_d)$ can be obtained analytically. Below we solve
these limiting cases.

\subsection{Large $v_{d}$}

First we consider the case of large $v_d$ which corresponds to large
parameters $d$. At $d\gg 1$ Eq.~\eqref{eq:1} gives $d=a^2\rho^2$.
Using it we find that the difference $p_{\rm{dc}}-\overline{p}
=(a-1)^2/2a^2$ at $v_d \gg 1$. Thus, at large $v_d$ the minimal power
is always smaller than $p_{\rm{dc}}$.

\subsection{Small $v_{d}$}\label{sec:smallvd}

We recall that in the rotational regime the motion is unbounded and
therefore $d> \rm{max}[U(\theta,\rho)]$, see Eq.~\eqref{eq:sol}.
This leads to $d>\rho^2$ for $\rho<1$ and $d>2\rho-1$ for $\rho>1$. We
show that it is possible to find analytical results for
$\overline{p}(v_d)$ when $d=\rho^2 +\epsilon$ or $d=2\rho-1+ \epsilon$
and $\epsilon\ll 1$.  Below we consider these two cases separately.

\subsubsection{$0<d<1$}

For the case $d=\rho^2 +\epsilon$ with $\epsilon\ll 1$,  we find
\begin{eqnarray}
T(\epsilon ,\rho )&=&\frac{1}{2}\int_{-\pi}^{\pi }
\frac{d\theta }{\sqrt{\epsilon +(\rho +\cos \theta)^{2}}}\label{eq:TsmallV}\\
T(\epsilon ,\rho )v_{d}&=&\pi a -\frac{1}{2}\int_{-\pi}^{\pi }
\frac{\cos \theta\,d\theta }{\sqrt{\epsilon 
+(\rho +\cos \theta)^{2}}}\label{eq:VsmallV}
\end{eqnarray}
In particular if $\rho =0$, one can find that $T =\ln (16/\epsilon )$,
$v_{d} =\pi a/\ln (16/\epsilon )$, and $\overline{p} =4/\ln
(16/\epsilon ) =4v_{d}/(\pi a)$.

From Eqs.~\eqref{eq:TsmallV} and \eqref{eq:VsmallV} we find
\begin{equation}
(v_{d}-\rho )T=\pi a -\frac{1}{2}\int_{-\pi}^{\pi}\frac{(\rho +\cos\theta)\, d\theta }{\sqrt{\epsilon +(\rho +\cos \theta)^{2}}}   
\label{eq:vTexact} 
\end{equation}
Also, substitution of $d=\rho^2 +\epsilon$ into Eq.~\eqref{eq:1} yields
\begin{equation}
\label{eq:rhoexact}
\int_{0}^{\pi}\!\! \sqrt{\epsilon +(\rho +\cos \theta)^{2}}\, d\theta =\pi a\rho,
\end{equation}
So far equations~\eqref{eq:vTexact} and \eqref{eq:rhoexact} are
exact. Note that the right-hand side of both equations
\eqref{eq:vTexact} and \eqref{eq:rhoexact} converge even if we set
$\epsilon =0$. Therefore, setting $\epsilon =0$ in them, we obtain
\begin{eqnarray*}
&&(v_{d}-\rho )T =a\pi -\frac{1}{2}\int_{-\pi}^{\pi}
\frac{\rho+\cos \theta }{|\rho +\cos \theta|} d\theta,
\\
&&\int_{0}^{\pi}|\rho +\cos\theta| d\theta = \pi a\rho,
\end{eqnarray*}
and finally
\begin{eqnarray}
&&(v_{d}-\rho )T = \pi (a+1) -2\theta_{\rho},
\\
&&2\rho \theta_{\rho} -\pi\rho +2\sin \theta_{\rho} = \pi a\rho,
\end{eqnarray}
where $\cos\theta_{\rho} =-\rho$.  This system of equations can be
rewritten as
\begin{eqnarray}
&&T = \frac{\pi (a+1) -2\theta_{\rho}}{v_{d}-\rho},
\label{eq:T1} \\
&&\tan\theta_{\rho} = \theta_{\rho}-\frac{\pi}{2} (1+a),
\label{eq:rho00}
\end{eqnarray}
and gives the period $T$ and parameter $\rho$ (since $\rho=
-\cos\theta_{\rho}$).  Since we look only for a solution of
Eq.~\eqref{eq:rho00} in the range $\pi /2<\theta_{\rho}<\pi$, we note
that this solution exists only for $a>1$.  Also, as $T$ must be
positive (otherwise there is no solution for $\epsilon$) we see that
$v_{d}>\rho$.

Thus, we conclude that at $a<1$ the minimal power for $v_d<1$ is
always achieved by dc current.  At $a>1$, based on Eqs.~\eqref{eq:T1}
and~\eqref{eq:rho00}, we find that $\rho =v_{\rm{rc}}$ for
$v_{d}>v_{\rm{rc}}$ (as long as $\epsilon$ is still small), and $\rho
=v_{d}$ for $0<v_{d}<v_{\rm{rc}}$. Therefore, for $v_{d}<v_{\rm{rc}}$,
$\overline{p}=v_{d}^{2}$ and period $T=\infty$ which corresponds to
the dc current case. For $v_{d}>v_{\rm{rc}}$,
\begin{equation}
\overline{p}=2v_{\rm{rc}}v_{d}-v_{\rm{rc}}^{2}, 
\end{equation}
and this power is minimized by the resonant ac current with the period
between picks given by
\begin{equation}
T=-\frac{2\tan \theta_{v_{\rm{rc}}}}{v_{d}-v_{\rm{rc}}}.
\end{equation}
In the limiting case of $a\gg 1$ we obtain $\rho\approx 2/a\pi \ll 1$,
and for $v_{d}\gg \rho$ we find $\overline{p}\approx 4v_{d}/a\pi $
with $T\approx a\pi /v_{d}$.

\subsubsection{$d>1$}

Now we consider the case of $d=2\rho-1+ \epsilon$ with $\epsilon\ll 1$
which corresponds to $v_{d}$ immediately above $v_{\rm{rc}}=1$ for
$a<1$. Neglecting very small $\epsilon$ the power becomes
\begin{equation}
\overline{p} (v_d)=1+2\rho_0 (v_d -1).
\label{eq:powersmalla}
\end{equation}
Note that the power $\overline{p}$ is linear in $v_d$ right above
$v_{\rm{rc}}=1$. Also for small $a$ one can see that at $v_{d}\gtrsim
1$ the power is significantly lower than $p_{\rm{dc}}$.

The parameter $\rho_0$ in Eq.~\eqref{eq:powersmalla} can be found from
Eq.~\eqref{eq:1} with $d= 2\rho -1$,
\begin{equation}
\int_{0}^{\pi} \sqrt{(1+\cos\theta)(2\rho_0 +\cos\theta-1)+\epsilon}\,\, 
d\theta =\pi a\rho_0.
\label{eq:dgr1}
\end{equation}
In the range where $\epsilon\ll 1$ we set it to zero and find from
Eq.~\eqref{eq:dgr1} the following equation for $\rho_0$:
\begin{equation}
\arcsin \frac{1}{\sqrt{\rho_0}} = \frac{\pi a}{2}
-\frac{\sqrt{\rho_0 -1}}{\rho_0}.
\end{equation}

\end{document}